\documentstyle[preprint,aps]{revtex}

\begin{document}
\draft
\tighten

\preprint{\vbox{\hbox{SUSX-TH-93/3/6} \hbox{IMPERIAL/TP/93-94/45}}}

\title{ Characteristics of Cosmic String Scaling Configurations }

\author{Daren Austin}
\address{School of Mathematical and Physical Sciences, University of
Sussex, \\Brighton BN1 9QH, United Kingdom}
\author{E.J. Copeland}
\address{School of Mathematical and Physical Sciences, University of
Sussex, \\Brighton BN1 9QH, United Kingdom and\\
Isaac Newton Institute for Mathematical Sciences, Cambridge CB3 0EH,
United Kingdom}
\author{T.W.B. Kibble}
\address{Blackett Laboratory, Imperial College, London SW7 2BZ,
United Kingdom and\\
Isaac Newton Institute for Mathematical Sciences, Cambridge CB3 0EH,
United Kingdom}

\date{} \maketitle

\begin{abstract}
Using the formalism developed in a previous paper we
analyze the  cosmological  implications of our conclusions
concerning the scaling behaviour of a network of cosmic strings, in
particular the idea that once gravitational back-reaction becomes
important the transient scaling  regime so far explored by numerical
simulations will be replaced by a new `full scaling' regime, with
slightly different parameters.  This has consequences for the
normalization of the string tension on all scales. We show how the new
scaling solutions affect the gravitational wave bound from
nucleosynthesis and the milli-second pulsar, and also describe the
consequences from large scale structure up to COBE scales.
\end{abstract}


\narrowtext

\noindent
 Cosmic strings formed at an early phase transition may
have interesting cosmological effects.  To test this idea, we need to
know how a network of strings evolves.   In a recent publication
\cite{ACK} we developed a detailed analytic formalism aimed at
improving our understanding of the evolution of such a network in
a $k=0$ Friedmann-Roberston-Walker (FRW) universe. The approach is
based on calculating the evolution of the probability, $p[{\bf
r}(l)]$, of a random segment of left-moving string of length $l$
having an extension ${\bf r}$ at time $t$.   One of our most
important conclusions was that the regime so far studied in numerical
simulations is a `transient scaling' regime, which will be succeeded
once gravitational back-reaction becomes important by `full scaling',
characterized by rather different parameters.   In this letter we
address some of the cosmological consequences arising from the new
results.

In \cite{ACK} we introduced three length scales $\xi(t),\bar{\xi}(t)$
and $\zeta(t)$ related to the long string density, the persistence
length along the long strings and the small-scale structure on the
string network respectively. In terms of the mean square extension,
$K(l) \equiv \overline{{\bf r}^2}$, $\bar\xi$ and $\zeta$ are defined
by
 \begin{eqnarray}
K(l,t) &\approx & 2\bar\xi(t) l, \qquad l \gg t,
 \label{Kbigl}\\
&\approx & l^2 - {l^3\over 3\zeta(t)}, \qquad l \ll t,
 \label{Ksmall}
 \end{eqnarray}
whereas $\xi(t) = (V/L)^{1/2}$ is the interstring
distance for a total  length, $L$, of string in a large volume $V$.
We introduced the dimensionless parameters $\gamma,\bar\gamma$ and
$\epsilon$ defined by   $\gamma = 1/H\xi$, $\bar\gamma = 1/H\bar\xi$,
$\epsilon = 1/H\zeta$,  where $H=\dot{R}/R$ is the Hubble parameter;
$R\propto t^{1/p}$, with  $p=2$ in the radiation era and $p={3\over2}$
in the matter era. With this definition, in  the radiation era,
$\gamma$ and $\bar\gamma$ are identical to the variables used in
\cite{KC} and \cite{CKA}, but in the matter era they are half as
large. In terms of these variables the evolution equations become
\cite{ACK}
 \begin{eqnarray}
-pt{\dot\gamma\over\gamma} &=& -p +  \left({3\over2}
-{\bar\alpha_{\rm nl}\over2} -
{F\over4}\right) +{c\over 2}\bar\gamma
+ {\Gamma G\mu\over2}\epsilon,
 \label{gamdot} \\
-pt{\dot{\bar\gamma}\over\bar\gamma} &=& -p+
\left(\bar\alpha_{\rm nl} + \bar{G} - {F\over2}\right) -
{\chi\gamma^2\over w\bar\gamma} + {I\over2}\bar\gamma + \Gamma
G\mu\epsilon,
 \label{gambardot} \\
-pt{\dot{\epsilon}\over\epsilon} &=& -p+
\left(\bar\alpha_{\rm nl}  + {3-12C\over2}F\right)
-{\chi\gamma^2\over\epsilon}
 \nonumber\\
&&+kc\bar\gamma +\Gamma
G\mu\hat C\epsilon.
 \label{epdot}
 \end{eqnarray}

The coefficient parameters and functions in
(\ref{gamdot})--(\ref{epdot}) were fully defined in \cite{ACK};
here we recall the definitions briefly.  The parameter $\bar\alpha =
1-2\langle\dot{\bf x}^2\rangle$ used in \cite{KC} and
\cite{CKA} was written as  $\bar\alpha = \bar\alpha_{\rm
nl}+{1\over2}F$, where $\bar\alpha_{\rm nl}$  relates to the effect of
loop formation and is probably small compared to $F$, which is defined
in \cite{ACK}, Eq. (7.17), and is the contribution
of stretching.  From the simulations of Bennett \& Bouchet
\cite{BB90}, we have, in the radiation- and matter-dominated eras
respectively,
 \begin{equation}
\bar\alpha_{\rm
RD} = 0.14\pm0.04 \qquad  \bar\alpha_{\rm MD} = 0.26\pm0.04,
 \label{alpha}
 \end{equation}
so $F$ is likely to be almost twice these
values.  The function $\bar G$, defined in \cite{ACK}, Eq. (7.23),
also represents a stretching effect; it
is larger than $F$ but of the same order, around unity.  The rate of
loop formation is determined by $c$, which probably lies in the range
0.1--0.5.  The effect of gravitational radiation is represented by the
parameter $\Gamma G\mu$, where $G$ is Newton's constant, $\mu$ the
string tension and $\Gamma$ a numerical constant of order 10--100.
The probability of intercommuting is governed by the constant $\chi$,
of order 0.1, and $w$, defined by the rate at which $K(l)$ approaches
its asymptotic form for large $l$, lies between 0 and 1.  The function
$I$ describes the effect of loop formation.  It is certainly less than
$2c$.  In \cite{ACK} it was argued that $I/2c$ is likely to be only
slightly less than unity, but this argument is not correct; in fact it
could in principle lie anywhere between zero and unity.
$C$ is a constant relating to the
stretching of strings on very small scales, which is probably also
small.  The two most critical parameters are $k$, which describes the
excess small-scale kinkiness on loops as compared to long strings, and
$\hat C$ which determines the rate at which gravitational
back-reaction smoothes the small-scale kinkiness.

Initially, soon after strings are formed, the gravitational radiation
terms in (\ref{gamdot})--(\ref{epdot}) are negligible.  Then,
as we argued in \cite{ACK}, $\gamma$ and $\bar\gamma$ will reach
scaling values, while (if $k$ is not too large --- see below)
$\epsilon$ starts to grow.  This is the {\it transient scaling\/}
regime.  The corresponding scaling values $\gamma_{\rm transient}$ and
$\bar\gamma_{\rm transient}$ are found by setting to zero the right
hand sides of (\ref{gamdot}) and (\ref{gambardot}).

This is the only scaling regime accessible to the numerical
simulations \cite{BB90,AT,BB88,AS90}.  From (\ref{gamdot}) and
(\ref{alpha}) above, assuming that $\bar\alpha_{\rm nl}$ is
negligible, we have (the error bars are our estimates):
 \begin{eqnarray}
c\bar\gamma_{\rm transient,RD} &= 1.14\pm0.04,
\nonumber\\
 c\bar\gamma_{\rm transient,MD} &= 0.26\pm0.04.
 \label{gabtran}
 \end{eqnarray}
The second of these is really a purely notional figure,
since the transient scaling regime will not last into the
matter-dominated era.   The simulations of \cite{BB90}  also give
 \begin{eqnarray}
\gamma_{\rm transient,RD} &= 7.2\pm1.4,
 \nonumber\\
\gamma_{\rm transient,MD} &= 2.8\pm0.7.
 \label{gamtran}
 \end{eqnarray}
With reasonable estimates for $c$, this shows that $\gamma_{\rm
transient}$ and $\bar\gamma_{\rm transient}$  are of roughly similar
magnitude.

In using Equations (\ref{gamdot})
and (\ref{gambardot}), which determine the
transient scaling values, we have assumed that $\bar\alpha_{\rm nl}$
and the gravitational radiation terms are negligible.  There are then
four remaining parameters: $\chi,w,c$ and $I$.  The latter two are
themselves in principle functions of $\gamma$ and $\bar\gamma$,
given by \cite{ACK}, Eqs.\ (6.17) and (6.27), but there are
considerable uncertainties in these formulae relating to the
effects of the small-scale structure, and it seems better therefore
at this stage to regard them as independent parameters; $c$ at least
can be estimated directly from the simulations.

We have explored numerically the solutions of Eqs.\
(\ref{gamdot}) and (\ref{gambardot}) for
the full ranges of the four parameters $\chi,w,c$ and
$\delta=1-I/2c$.  Both $w$ and $\delta$ are restricted to lying within
the range 0 to 1.  In general, we find that the value of $c$ is
fairly tightly constrained; the other parameters less so.  In the
radiation era, for $\chi=0.1$, the ranges within which we find
solutions consistent with Eqs.\ (\ref{alpha})--(\ref{gamtran}) are
 \begin{eqnarray}
0.10<&c&<0.18,\nonumber\\
0.62<&w&<1,\qquad{\rm (RD)} \\
0<&\delta &<0.21.\nonumber
 \end{eqnarray}
Generally speaking, the smaller values of $w$ or the larger values of
$c$ require small values of $\delta$.  In the matter era, conditions
are less restrictive; in fact $\delta$ is unconstrained:
 \begin{eqnarray}
0.08<&c&<0.14,\nonumber\\
0.26<&w&<1,\qquad{\rm (MD)} \\
0<&\delta &<1.\nonumber
 \end{eqnarray}
In both cases, lowering the value of $\chi$ tends to relax the
constraints; with $\chi=0.06$, much wider ranges of $w$ and $\delta$
are possible --- for example, in the radiation era $\delta$ can be as
large as 0.6.

The predicted values of $\gamma$ are generally towards the lower end
of the range allowed by (\ref{gamtran}):
 \begin{eqnarray}
5.8<\gamma_{\rm transient,RD}<6.5,&\quad&
6.5<\bar\gamma_{\rm transient,RD}<11,
 \nonumber\\
2.1<\gamma_{\rm transient,MD}<2.8,&\quad&
1.7<\bar\gamma_{\rm transient,MD}<3.0.
\label{tranrange}
 \end{eqnarray}
It is interesting that $\bar\gamma$ tends to be
bigger than $\gamma$ in the
radiation era, but about the same size in the matter era.  We also note
that $\chi$ and $\gamma$ always appear in
(\ref{gamdot})--(\ref{epdot}) in the combination $\chi\gamma^2$; hence
lowering $\chi$ will always increase the scaling value of $\gamma$.

During the transient scaling regime, $\epsilon$ grows until the point
where the gravitational radiation terms become important.  The
important question is whether then a {\it full scaling\/} regime is
reached, in which all three of $\gamma,\bar\gamma,\epsilon$ are
constant.  If there is such a solution, it is in the region of
parameter space where $\epsilon\gg\gamma_{\rm full}\sim\bar\gamma_{\rm
full}$, which means that the intercommuting term in the third
equation, $\chi\gamma^2/\epsilon$, is negligible.  If we neglect it,
 then, assuming we know the values of the various constants,  it
is straightforward to solve for the scaling parameters.  As we showed
in \cite{ACK}, the existence of a full scaling solution depends
primarily on the magnitudes of $\hat C$ and $k$; we require $\hat C >
\hat C_{\rm cr} > k$,
where
\begin{equation}
\hat C_{\rm cr} = {p - \bar{\alpha}_{\rm nl} - \frac{3}{2} (1 - 4C) F
\over 2p - 3 + \bar{\alpha}_{\rm nl} + \frac{1}{2} F}.
\label{chcrit}
\end{equation}
and is typically of order one. Unfortunately, we do not know the
values of $\hat C$ and $k$, but we shall assume, as seems likely,
that these inequalities are satisfied.

It is important to realize that the simulations performed so far,
which neglect gravitational radiation, may have provided misleading
information about the true values of the scaling  parameters.  From
the evolution equations, (\ref{gamdot})--(\ref{epdot}), it is
clear that when the gravitational radiation terms become significant,
then $\bar\gamma$ decreases. Without knowing the values of $\hat
C$ and $k$ it is not possible to predict the extent of the decrease,
but what we can do is to examine the relationships between the
changing values of $\gamma,\bar\gamma$ and $\epsilon$.

 From Eq.\ (\ref{gamdot}), we might expect that when full scaling is
reached the values of $c\bar\gamma$ and of $\bar\epsilon=\Gamma
G\mu\epsilon$ should be comparable; as $\bar\epsilon$ increases,
$c\bar\gamma$ must decrease.  The ratio between the two is determined
by the values of $\hat C$ and $k$; by \cite{ACK}, Eq.\ (9.22),
 \begin{equation}
{\bar\epsilon\over c\bar\gamma} \equiv
{\Gamma G\mu\epsilon\over c\bar\gamma} =
{\hat C_{\rm cr}-k\over\hat C-\hat C_{\rm cr}}.
\label{ratio}
 \end{equation}
We have studied the effect of changing $\bar\epsilon$ on the values
of $\gamma$ and $\bar\gamma$ predicted by Eqs.\ (\ref{gamdot}) and
(\ref{gambardot}).  As we noted, as $\bar\epsilon$ increases,
$c\bar\gamma$ decreases.  It is not so obvious from the
equations what happens to $\gamma$. However, we find that in all cases
it too decreases, though by a smaller factor.

In the radiation era, the dependence of $\gamma$ and $\bar\gamma$ on
$\bar\epsilon$ is nearly linear; less so in the matter era.  For
example, if we choose the parameter values $\chi=0.1, c=0.1,
\delta=0.1, w=0.8$, we find the dependence shown in Figure 1.  Other
parameter values within the allowed range give very similar curves.
When $\bar\epsilon=c\bar\gamma$, $\bar\gamma$ has decreased by nearly a
half in the radiation era, and in the matter era by about a third.  At
the same point, $\gamma$ has decreased in the radiation era by about
25\%; in the matter era by 10\%.

 From (\ref{tranrange}), reasonable estimates seem to be
 \begin{eqnarray}
\gamma_{\rm full,RD} = 3\;{\rm to}\;6,&\qquad&
c\bar\gamma_{\rm full,RD} = 0.2\;{\rm to}\;1,
\nonumber\\
\gamma_{\rm full,MD} = 1.4\;{\rm to}\;2.6,&\qquad&
c\bar\gamma_{\rm full,MD} = 0.05\;{\rm to}\;0.27,
 \label{gamful}
 \end{eqnarray}
For observational tests, we need to know how far out we should expect
to have to look to find a cosmic string.  What is most important here is
the scaling value of $\gamma$ in the recent, matter-dominated epoch.
With $H_0=2/3t=100h\; {\rm km\;s^{-1}Mpc^{-1}}$ ($0.5<h<1$), the mean
interstring distance today would be $(\gamma_{\rm
full,MD}H_0)^{-1} = (1200\;{\rm to}\;2000)  h^{-1}$ Mpc, with the
nearest string to us being a distance say half that, {\it i.e.}\ in the
range $z = 0.2\;{\rm to}\; 0.5$.

We are also of course interested in the nearest loops, so we turn
next to them.  The loops that are  born off the network are usually
produced with quite  large centre of mass velocities, and lose energy
both through  red-shifting and, at a constant rate, to  gravitational
radiation,   $\dot E_{\rm gr} = - \Gamma_{\rm loops} G \mu^2$,
until their rest-mass is reduced to zero.  Here $\Gamma_{\rm loops}$
is a constant of order 50--100.   We assume that  the fraction of the
total  energy which will eventually be radiated is $f$, {\it i.e.},
$(1-f)E_{\rm b}$ is red-shifted away, where   $E_{\rm b}$ is the
average energy once it has reached a stable  non-self-intercommuting
form.

As a first approximation, we imagine that the loop  immediately loses
the fraction $(1-f)$ of its total energy, leaving  a static  loop of
length $fE_{\rm b}/\mu$, which loses its energy by  gravitational
radiation, a fairly good picture as long  as the  lifetime is long
compared to $t_{\rm b}$. Given this, we take the effective  length of
the loop at birth to be  $l_{\rm b} = fE_{\rm b}/\mu \equiv
(\kappa-1)\Gamma_{\rm loops} G \mu t_{\rm b}$,  where we have
introduced the  dimensionless  parameter $\kappa$. Note that $\kappa$
is identical to the $1/\beta$  of Allen and  Caldwell \cite{CA92} and
Allen and Shellard \cite{AS92}. At  a later time the loop's
length is  $l = \Gamma_{\rm loops} G\mu(\kappa t_{\rm b} -t)$;  it
disappears at $t_{\rm d} = \kappa t_{\rm b}$. Numerical  simulations
indicate $\kappa$ to be somewhere between 2 and 10, the latter value
certainly satisfying the asumption we have made on the role of the
initial red shifting.   (Even for $\kappa\approx2$, the corrections
are not large.)

In a volume $V$, the rate at which loops are born from long strings
is given by,
 \begin{equation}
\dot{N} = {\nu V \over (\kappa-1)\Gamma_{\rm loops} G\mu t_{\rm b}^4},
 \label{ndot}
 \end{equation}
 where
 \begin{equation}
\nu ={fc\bar\gamma\gamma^2
\over p^3}.
 \label{nu}
 \end{equation}
Taking $f$ to be  around  0.7  while  $\gamma$ and $c\bar\gamma$
are given by the estimates above, we obtain
 \begin{equation}
\nu_{\rm RD} = 0.2\;{\rm to}\;3, \qquad \nu_{\rm MD} = 0.02\;{\rm
to}\;0.4.
 \label{newnu}
 \end{equation}
The parameter $\nu$ is not very well constrained.  However, we
predict a  somewhat lower rate of loop formation than for example ref.\
\cite{CA92}.

Any possible direct observational test of the cosmic string
scenario  depends on knowing how many long strings or loops we may
expect to see  within a reasonable distance, and how large they would
be.  We  therefore turn to estimates of these quantities.

The total number density of loops at time $t$ is obtained from
$\dot{N}$,  allowing for the expansion factor:
 \begin{equation}
\hat{n}=\int_{t/\kappa}^{t} {\dot{N}(t_{\rm b}) \over V} dt_{\rm b}
\left({R_{\rm b} \over R}\right)^3,
 \label{ntotal}
 \end{equation}
where $R(t)$ is the FRW scale factor.  Except for the
transition region between RD and MD, we obtain
 \begin{equation}
\hat{n} = A(\kappa){\nu \over \Gamma_{\rm loops} G\mu
t^3},
 \end{equation}
 with
 \begin{equation} A_{\rm RD}(\kappa) =
{2\over3}{\kappa^{3/2}-1\over \kappa-1},
 \qquad A_{\rm MD}(\kappa) = 1.
 \label{aofk}
 \end{equation}

This then gives the mean distance between loops, the present distance
being
 \begin{equation}
\hat{n}^{-1/3} = \left({\Gamma_{\rm loops}
G\mu \over \nu}\right)^{1/3} t  = \left({\Gamma_2 \mu_6 \over
\nu}\right)^{1/3} 90 h^{-1}~{\rm Mpc}
 \label{meandist}
 \end{equation}
 where  $\Gamma_2 = \Gamma_{\rm loops}/100$  and $\mu_6 = 10^6
G\mu$. Taking   $\nu_{\rm MD} = 0.1$,  $\Gamma_{\rm loops} = 100$,
$G\mu=10^{-6}$, we obtain  $\hat{n}^{-1/3}  \simeq 200 h^{-1}~{\rm
Mpc}$.  The nearest loop to us is probably at about half this
distance.  If this is the case there are probably a few hundred loops
with  $z<0.3$.

The total length of all loops in a volume $V$  at time $t$ is given by
 \begin{equation}
L_{\rm loops} = \int_{t/\kappa}^{t}
{\dot{N}(t_{\rm b}) \over V} dt_{\rm b} \left({R_{\rm b} \over
R}\right)^3 \Gamma_{\rm loops} G\mu (\kappa t_{\rm b} - t),
 \label{ltotal}
 \end{equation}
 from which we can write the energy
density in loops,
 \begin{equation}
 \rho_{\rm loops} = B(\kappa){\mu
\nu \over t^2},
 \end{equation}
where
 \begin{eqnarray}
B_{\rm RD}(\kappa) &= {2(2\kappa^{3/2} -3\kappa +1) \over
3(\kappa-1)},
 \nonumber\\
B_{\rm MD}(\kappa) &= {\kappa\over \kappa-1}\ln \kappa -1.
 \end{eqnarray}
The mean length of a loop is
 \begin{equation}
\bar{l} = {B(\kappa)\over A(\kappa)}\Gamma_{\rm loops} G\mu t.
 \label{meanl}
 \end{equation}
 Thus at the present time, we have
 \begin{equation}
\bar{l} = B_{\rm MD}(\kappa)\Gamma_2 \mu_6
200h^{-1}\;{\rm kpc},
 \label{ltoday}
 \end{equation}
with a corresponding mass
 \begin{equation}
M_{\rm loop} = \mu \bar{l} =
B_{\rm MD}(\kappa)\Gamma_2 \mu_6^2 4\times10^{12} M_{\odot}.
 \label{mtoday}
 \end{equation}
The median loop size is given by the
same expression but with  $B_{\rm MD}$ replaced by
${\kappa-1\over\kappa+1}$.  Thus we see  that  the size of a typical
loop is larger than a typical  galaxy while its mass is that of a very
large galaxy.  In other words, surviving loops are {\it significant}
objects gravitationally.

The total gravitational wave  energy generated by cosmic strings is
the sum of that generated by loops and by long strings.  In the RD
era, we have
 \begin{equation}
\rho_{\rm gr}(t) = \mu[A_{\rm RD}(\kappa)\nu + \nu_{\rm ls}]{1\over
t^2}   \int_{t_{\rm f}}^{t} {dt_{r}\over t} \left({t\over t_{\rm r}}
\right)^3\left({R_{\rm r} \over R}\right)^4,
 \end{equation}
where
 \begin{equation}
\nu_{\rm ls} = {\Gamma G\mu\epsilon\gamma^2 \over p^3}.
 \end{equation}

Here  the lower limit is the time of formation of cosmic strings, or
rather the time at which they  start to move freely and so lose a
significant amount of energy to  gravitational radiation. It is
interesting to note that the ratio between the parameters $\nu_{\rm
ls}$ and $\nu$ which govern the contributions of long strings and
loops is given by
 \begin{equation}
{\nu_{\rm ls}\over\nu} = {\Gamma G\mu\epsilon\over fc\bar\gamma},
 \end{equation}
  {\it i.e.}, essentially the ratio (\ref{ratio}).  Since
$\bar\gamma$ decreases as $\epsilon$ increases, if $\nu_{\rm ls}$ is
near its upper limit, then $\nu$ must be relatively small, and vice
versa --- if more energy is dissipated as gravitational radiation less
must be available for loop formation.

The spectrum of gravitational waves emitted by cosmic strings  is
constrained by the nucleosynthesis limit on relativistic particle
species,  as well as the  pulsar timing on a stochastic gravitational
wave background. At the time of nucleosynthesis we obtain
 \begin{equation}
\Omega_{\rm gr}(t_{\rm nuc}) = {32 \pi G\mu\over 3}
[A_{\rm RD}(\kappa)\nu + \nu_{\rm ls}] \left({g_{*,{\rm nuc}}\over
g_{\rm *,r}}\right)^{1/3}  \ln {t_{\rm nuc} \over t_{\rm f}}.
 \label{grnuc}
 \end{equation}
Typically, we have  $T_{\rm nuc}/T_{\rm
f}  \simeq 10^{-16}$, $g_{*,{\rm nuc}}/g_{\rm *,r} = 10.75/106.75
\simeq 0.1$.  With these values, (\ref{grnuc}) becomes
 \begin{equation}
\Omega_{\rm gr}(t_{\rm nuc}) =
1.2\times10^{-3} \mu_6  [A_{\rm RD}(\kappa)\nu + \nu_{\rm ls}].
 \label{nucbound1}
 \end{equation}

The nucleosynthesis limit places a limit on the energy density  in
neutrinos plus any relativistic particle species beyond the photons
and  electrons  present at $t_{\rm nuc}$. It is bounded by the
equivalent energy  density of $N_{\nu}$ neutrino species. In terms of
the energy density in  gravitational radiation at nucleosynthesis we
obtain
 \begin{equation}
\Omega_{\rm gr} \le 0.163 \times (N_{\nu} -3)
\Omega_{\rm rad},
 \label{nucbound}
 \end{equation}
where  three  light neutrino species have been assumed.
 According to \cite{olive94},  the current
observational limit on light ($\le 10$MeV) neutrino species
is $N_{\nu} \le 3.1$, which yields the bound
 \begin{equation}
\Omega_{\rm gr}(t_{\rm nuc}) \le 0.02.
 \label{nucbound2}
 \end{equation}
Combining the two expressions we obtain
 \begin{equation}
\mu_6[A_{\rm RD}(\kappa) \nu + \nu_{\rm ls}] < 15.
 \label{weak}
 \end{equation}
which is not at all restrictive for any values of $\nu$ and $\kappa$ in the
expected range.
 A more stringent limit, $N_{\nu} \le 3.04$, has been given
recently  \cite{KK94}.  Using this limit  would
change the bounds in (\ref{nucbound2}) and (\ref{weak}) to 0.007 and
5, respectively.  This bound is still not hard to satisfy, but may
be somewhat restrictive if either of the terms within the square
brackets in (\ref{weak}) is close to its upper limit.

This though is probably not the tightest gravitational  radiation
bound. The pulsar timing limit requires  that the energy density in
gravitational  radiation, in a logarithmic interval at  $f =(8.2~{\rm
years})^{-1}$ satisfy the inequality \cite{ryba91}
 \begin{equation}
{d\Omega_{\rm gr} \over d \ln f}\Bigg|_{f =(8.2~{\rm years})^{-1}}  \le
1.0 \times 10^{-7} h^{-2}.
 \label{pulsar}
 \end{equation}
This gives
 \begin{equation}
\mu_6 <
0.2\sqrt{\kappa-1\over\kappa}\left(10\over\kappa\right)
\left(0.1\over\nu\right)^{3/2} {\Gamma_2^{1/2}\over h^{7/2}}.
 \end{equation}
 (The details of this derivation will be presented
elsewhere.)   Note the strong dependence on $\nu$, $\kappa$ and $h$.
For $h=0.5$ the condition is not particularly restrictive, though it
puts bounds on $\nu$  and $\kappa$, but for $h=1.0$ it is a very
restrictive condition, which could only be satisfied by making $\kappa$
and $\nu$ very small, say $\kappa =2$, $\nu\le 0.05$.

The growth of primordial density perturbations generated by strings
provides a separate constraint on $G\mu$. In the work of Albrecht and
Stebbins \cite{ASt92} with strings and hot dark matter,  the power
spectrum of the overdensity fluctuations is approximated by an
integral in conformal time over the product of the  transfer function
describing how the initial perturbation spectrum is evolved to
the present, and a ``form factor'', ${\cal F}$, which describes
the  input from the string. It is in this term that the string scaling
behaviour is hidden.  In particular ${\cal F}$ is given in terms of the
wavenumber $k$ as
 \begin{equation}
{\cal F}(k\xi/R) = {2\over \pi^2} \overline
{\beta^2 \Sigma}{\chi^2\over \xi^2} \left({1\over
1+2(k\chi/R)^2}\right).
 \label{formfac}
 \end{equation}
Briefly, $\xi$ and  $\chi$ correspond roughly to $\xi$ and
$\bar\xi$ in our notation,  $\beta$ gives the r.m.s.\ velocity of the
strings, and $\Sigma$ the increase in the surface density of the wakes
due to the wiggliness.  For simplicity,  we will assume  that the
only parameters that are affected are $\chi$ and $\xi$. Albrecht and
Stebbins incorporate the small scale-structure on  the network into a
renormalized string tension $\mu$, and let this value vary  depending
on the  type of string they are attempting to mimic. We would like to
know in what direction the power spectrum goes if the scaling
solutions are  changed, by increasing $\chi$ and $\xi$.  If $\chi/\xi$
is doubled (which is  roughly the upper limit of  what we might
expect), then the form factor $\cal F$ is multiplied by 4 and so
therefore is the power spectrum.  Hence in order to fit the  large
scale structure data, we would naively expect that the  normalization
of $\mu$ must go down by a factor of two.  For the case where  they
attempt to match the simulations of ref.\ \cite{BB90}, they use
$\overline{\beta^2 \Sigma} \sim 1.2$ and (for $h=0.5$) $G\mu_{\rm ren}
= 4.0 \times 10^{-6}$. We would obtain $G\mu_{\rm ren} \approx 2
\times 10^{-6}$, a value well within current allowed limits on $G\mu$.

The third observational test which constrains the string parameters
arises from an\-iso\-tropy observations on the microwave background
sky  on both large and small angular  scales. There are few cosmic
string calculations of this  effect to date, unlike say the situation
in inflation or global textures. In a recent paper, Perivolaropoulos
\cite{leandros93} obtains an analytic expression for $\Delta T/T$ in
terms of the interstring distance $\xi$. There are many assumptions
in the calculation,  including the use of straight strings and the
neglect of loops, which make direct comparison somewhat dubious.
However, his final result, in our notation, is that $\Delta T/T
\propto G\mu \gamma$.   Perivolaropoulos quotes a result of $G\mu =
(2.0 \pm 0.5) \times 10^{-6}$.  If as we believe, $\gamma$ is somewhat
less than the prediction of ref.\ \cite{BB90}, this should be
increased,  but only very slightly.

We conclude that the modifications suggested by our analytic study
slightly weaken the observational constraints on $G\mu$.  For $h=0.5$,
they are fully compatible with current observational limits on this
parameter, but for $h=1.0$, it is very difficult to accommodate the
gravitational-wave bound.  However, there are still many uncertainties
that need to be resolved before we can make really definitive
predictions.  The biggest problem is to find a way of estimating the
parameters associated with the small-scale structure on the strings,
in particular $\hat C$ and $k$  which determine the effect on the
small scale length of the gravitational back-reaction and loop
formation, respectively.

\acknowledgments

We have benefited from conversations with Andreas Albrecht.  D.A.\ and
E.C.\ are indebted to PPARC for support.  We are grateful to Robert
Caldwell and to the referee for pointing out errors in the original
version of the paper.  E.C.\ and T.K.\ are happy to acknowledge the
hospitality of the Isaac Newton Institute for Mathematical Sciences,
Cambridge, where the work was completed.

\begin{figure}

\caption{}{Variation of $\gamma$ and $\bar\gamma$ with $\bar\epsilon$ in
the full scaling regime ({\it a}) in the radiation-dominated and ({\it b}) in
the matter-dominated era.}

\end{figure}

\end{document}